\documentclass[showpacs,preprintnumbers,amsmath,amssymb,nofootinbib]{revtex4}

\usepackage{graphicx}
\usepackage{dcolumn}
\usepackage{bm}
\usepackage{latexsym}
\usepackage{epsfig}
\usepackage{pbox}
\usepackage{multirow}
\usepackage{rotating}
\usepackage{bm}

\newcommand{\be}{\begin{equation}}
\newcommand{\ee}{\end{equation}}
\newcommand{\beqq}{\setlength\arraycolsep{2pt}\begin{eqnarray}}
\newcommand{\eeqq}{\vspace{0cm} \end{eqnarray}}
\newcommand{\bea}{\begin{eqnarray}}
\newcommand{\eea}{\end{eqnarray}}

\begin{document}

\title{$\Lambda(t)$ cosmology induced by a slowly varying Elko field}

\author{S. H. Pereira} \email{shpereira@feg.unesp.br}
\author{A. Pinho S. S.} \email{alexandre.pinho510@gmail.com}
\author{J. M. Hoff da Silva} \email{hoff@feg.unesp.br}
\affiliation{Universidade Estadual Paulista (Unesp)\\Faculdade de Engenharia, Guaratinguet\'a \\ Departamento de F\'isica e Qu\'imica\\ Av. Dr. Ariberto Pereira da Cunha 333\\
12516-410 -- Guaratinguet\'a, SP, Brazil}

\author{J. F. Jesus} \email{jfjesus@itapeva.unesp.br}
\affiliation{Universidade Estadual Paulista (Unesp)\\ Campus Experimental de Itapeva \\ R. Geraldo Alckmin, 519\\ Itapeva, SP, Brazil.}


\begin{abstract}
In this work the exact Friedmann-Robertson-Walker equations for an Elko spinor field coupled to gravity in an Einstein-Cartan framework are presented. The torsion functions coupling the Elko field spin-connection to gravity can be exactly solved and the FRW equations for the system assume a relatively simple form. In the limit of a slowly varying Elko spinor field there is a relevant contribution to the field equations acting exactly as a time varying cosmological model $\Lambda(t)=\Lambda_*+3\beta H^2$, where $\Lambda_*$ and $\beta$ are constants. Observational data using distance luminosity from magnitudes of supernovae constraint the parameters $\Omega_m$ and $\beta$, which leads to a lower limit to the Elko mass. Such model mimics, then, the effects of a dark energy fluid, here sourced by the Elko spinor field. The density perturbations in the linear regime were also studied in the pseudo-Newtonian formalism.
\end{abstract}

\maketitle


\section{Introduction}

The $\Lambda$CDM model, known as the standard model of cosmology, is constructed with the assumption that the cosmological constant term, $\Lambda$, dominates the recent accelerated evolution of the universe, along with a dark component of Cold Dark Matter (CDM) and the baryonic visible matter. The dark matter (DM) component obviously does not interact electromagnetically with the baryonic matter. Such model is in good agreement with recent observational data \cite{SN,SuzukiEtAl12,wmap,planck}. Although being the best model in explaining the observational data, some concerns about its origin and the great discrepancy between the theoretical and the expected value of the cosmological constant term are yet open questions \cite{CC}. Even other important issue is why does we live in a special phase of the cosmic evolution where the contribution of $\Lambda$ is of the same order of magnitude of the baryonic and dark matter components. This is known as \emph{cosmic coincidence problem}. 

In the same line of models based on a cosmological term, more general scenarios considering a time varying cosmic term have been recently proposed in the literature. The main aim of these models is to explain the effect of recent acceleration of the universe \cite{lambdaDec,termoLdec,carneiroLdec,sandro}. Such models are called $\Lambda(t)$ cosmology or vacuum decay cosmology, but its theoretical motivation and nature also constitutes an open question. Part of these works are motivated by corrections to the vacuum energy in theories of renormalization group \cite{ren}. According to the general framework, the vacuum quantum fluctuations are the source of a vacuum energy density that depends on the space-time curvature. After a renormalization procedure, the divergent vacuum contribution is subtracted and the resulting effective energy density is allowed to decay into DM particles \citep{termoLdec,carneiroLdec}.

Models in which the cosmological term are absent but its effects are mimicked by an exotic fluid of negative pressure filling the whole universe are called Dark Energy (DE) models. Several models of DE have been studied recently \cite{reviewDE,DER,scalar,DEspinor,DE1,DE3,abwa,Ioav}, including models of scalar field, exotic fluids or interacting DE - DM models. In addition to the models that use standard scalar fields as DE or DM candidates, recent works have shown that there is a new kind of non-standard spinor with interesting ``dark'' properties which could be useful in order to describe both DM as DE. It is named Elko (\textit{Eigenspinoren des ladungskonjugationsoperators}) and was proposed by by Ahluwalia-Khalilova and Grumiller \cite{AHL1,AHL2,AHL3} (see next section for more details). After its discovery, several theoretical aspects of the Elko spinor field has been studied \cite{roldao1,julio1} and also some interesting cosmological applications \cite{FABBRI,BOE6,BASAK,BOE7,WEI,sadja,kouwn,saj,js,asf}. The searching for such kind of particle at CERN LHC has also been addressed recently \cite{dias1,alves1}. 

Since the Elko spinor is a genuine fermion, it is possible to envisage its coupling also to torsion fields, investigating its cosmological consequences \cite{FABBRI,kouwn}. In this paper we present the complete equations of a Elko spinor field coupled to gravity in an Einstein-Cartan framework and show that a slowly varying Elko spinor field presents the same behavior of a time varying cosmic term, which leads to a dark energy evolution. As we have already emphasized, varying cosmic term models may be understood as a sort of {\it ad hoc} approaches to the cosmic evolution problem. Therefore, it is indeed relevant to obtain its cosmological behavior from the field theory realm. Moreover, we show that the positivity of the Elko mass leads to a nearly null self-coupling for quartic vertex.

This paper is organized as follows: Section II briefly presents the main properties of the Elko spinor field; Section III, together with the Appendix, present the equations of the Elko field into an Einstein-Cartan framework; Section IV concerns the slowly varying Elko spinor field; Section V presents the observational constraints from SN Ia and BAO; Section VI presents a study of linear evolution of density perturbations and conclusions are left to Section VII.


\section{Elko spinor fields}

More or less recently, a new class of spinor fields have been proposed in the literature \cite{AHL1,AHL2,AHL3}, with several interesting properties, both from theoretical and phenomenological reasons. Such new class of spinors, the so-called Elko spinors, in contrast to the Dirac spinor field, perform a complete set of eigenspinors of the charge conjugate operator, endowing the spinor field of darkness, i. e., the spinors are neutral with respect to gauge interactions. Moreover, these spinors are shown to have canonical mass dimension one, instead of three-half. All together, the field construction reveals a quite interesting DM or even DE candidate. As an aside (but important) remark, we stress that the quantum field theory no-go theorems \cite{weinb} are not applicable to the case in question since the field violate the full Lorentz symmetry in a subtle way. Actually, the situation can be better explained by making clear some of the main steps of the field construction. In this section we shall pinpoint these steps in an argumentative ground. For a comprehensive and up to date review, please see Ref. \cite{novodual}.   

After the formal development of the Elko spinors and the treatment of its basic properties, it is possible to envisage that the theory shows the existence of a preferred direction, jeopardizing the full relativistic symmetry of the theory. In fact, the naive use of the Dirac adjoint leads to a vanishing norm. There are, however, a judicious construction fully based on relativistic argues which arrives at the right dual to the Elko case \cite{dual}. We mention this last fact because there is an important fingerprint associated to the new dual: the spin sums carries the Lorentz broken term, instead of some different term in the dispersion relation. The associated quantum field inherits this property rendering non-locality, except along a preferred direction\footnote{There is a recent further development in the dual construction evading the Lorentz violation in the spin sums \cite{novodual,nois,teorema}, but keeping all the other characteristics of the field. Since we remain the scope of this work in the cosmological consequences realm, most of the results we report are still valid even in this last case.}. Nevertheless, it was demonstrated that the violation can be recast into subgroups of the Lorentz group, just as the groups $SIM(2)$ and $HOM(2)$ \cite{PC}. These subgroups can be obtained from the full Lorentz group by removing the discrete symmetry operators and rearranging the remain generators \cite{Aluva}. 

The consequences to the cosmological set up we shall investigate in this work are analyzed bearing in mind a classical spinorial field, taking advantage of previous results \cite{js}, as the time factorization of the spinor field possibility. In order to accomplish our task, it is important to emphasize once again that the spinor field satisfies only the Kelin-Gordon equation: the quantum field has a scalar-like propagator, whilst the classical spinor is not annihilated by the Dirac operator. In this vein the lagragian to be associated to the theory has the kinetic term given by $(\nabla_{\mu}\lambda)^2$. In the following we shall investigate the Elko spinor field coupled to gravity in a Einstein-Cartan framework, with a Friedmann-Robertson-Walker metric and all the particularities coming from these spacetime specifications are to be encoded into the covariant derivative. 

We finalize this  section by reinforcing the spinorial character of Elko fields, this time calling attention to its (while a fermion) coupling to torsion fields. As we are going to see, the cosmological impact of such possibility is indeed worth to be reported. Beyond that, it is indeed a matter of subtlety the fact that the (not yet experimentally verified) torsion fields may be associated to Elko dark spinors contributing to describe the behavior of DE.


\section{Elko cosmology in an Einstein-Cartan framework}

We start introducing the Elko spinor action in a general Einstein-Cartan framework:
\begin{equation}
S = \int d^4 x \sqrt{-g} \left[ -\frac{1}{2\kappa^2}\tilde{R}+{1\over 2}g^{\mu\nu}\tilde{\nabla}_\mu \bar{\lambda}\tilde{\nabla}_\nu \lambda -V(\bar{\lambda}\lambda) \right] \,,
\label{actionE}
\end{equation}
where $\kappa^2\equiv8\pi G=1/M^2_{pl}$ with $c=1$. The tilde denotes the presence of torsion terms into the Ricci scalar $\tilde{R}$ and covariant derivatives, namely, $\tilde{\nabla}_\mu \lambda\equiv \partial_\mu\lambda - \Gamma_\mu \lambda$ and $\tilde{\nabla}_\mu \bar{\lambda}\equiv \partial_\mu\bar{\lambda} + \bar{\lambda}\Gamma_\mu$,
where $\Gamma_\mu$ is the connection associated to spinor fields, containing the spin connection $\omega_\mu^{\;\;a b}$. We leave for the Appendix a detailed derivation of the basic stuff concerning the field equations, since such program was already carried out by Kouwn et al \citep{kouwn}. From now on, the non-vanishing torsion terms are introduced via the $h(t)$ and $f(t)$ functions. The Elko spinor fields are assumed as $\lambda(x^\mu)=\phi(t) \xi({\bf x})$ and $\bar{\lambda}(x^\mu)=\phi(t) \bar{\xi}({\bf x})$. Besides, the normalization $\bar{\xi}\xi =1$ will be used without lost of generality. 

The flat Friedmann-Robertson-Walker (FRW) metric reads $ds^2=N(t)^2dt^2-a(t)^2(dx^2+dy^2+dz^2)$, where $N(t)$ is a lapse function and $a(t)$ is the scale factor of the universe. The lagrangian density that follows from (\ref{actionE}) is
\begin{equation}
\mathcal{L} = -{1\over N}\bigg({3a\dot{a}^2\over \kappa^2} - {3a^3h^2\over \kappa^2} - {1\over 2}a^3\dot{\phi}^2-{3\over 8}a(\dot{a}+ah)^2\phi^2\bigg)-N\bigg({3a^3f^2\over \kappa^2}+{3\over 8}a^3f^2\phi^2+a^3V(\phi)\bigg)\,,\label{lagran}
\end{equation}
where $V(\phi)$ is the potential.

The Friedmann equations and the equation of motion for $\phi(t)$ can be obtained by studying the Euler-Lagrange equations coming from (\ref{lagran}) with respect to $N(t)$, $a(t)$ and $\phi(t)$, respectively. First, in order to eliminate the functions encoding the torsion contribution, $h(t)$ and $f(t)$, we investigate its equations of motion
\begin{equation}
h(t)=-{1\over 8}{\kappa^2\phi^2\over (1+\kappa^2\phi^2/8)}\bigg({\dot{a}\over a}\bigg)\;, \qquad f(t) = 0\,.\label{hf}
\end{equation}
Now, the Friedmann equation for the energy density can be obtained by taken the Euler-Lagrange equation with respect to the lapse function $N(t)$ (setting $N\to 1$), and substituting $f(t)$ and $h(t)$ from (\ref{hf}):
\begin{equation}
H^2={\kappa^2\over 3}\bigg[{\dot{\phi}^2\over 2}+V(\phi)+{3\over 8}{H^2\phi^2\over (1+\kappa^2\phi^2/8)} \bigg]\,,\label{H2}
\end{equation}
where $H=\dot{a}/a$. In the same fashion, the pressure equation is obtained by taken the Euler-Lagrange equation with respect to $a(t)$:
\begin{equation}
-2\dot{H}-3H^2={\kappa^2}\bigg[{\dot{\phi}^2\over 2}-V(\phi)-{3\over 8}{H^2\phi^2\over (1+\kappa^2\phi^2/8)} - {1\over 4 }{\dot{H}\phi^2\over (1+\kappa^2\phi^2/8)}-{1\over 2}{H\phi\dot{\phi}\over (1+\kappa^2\phi^2/8)^2} \bigg]\,.\label{Hdot}
\end{equation}
Finally, the equation of motion for $\phi(t)$ reads
\begin{equation}
\ddot{\phi}+3H\dot{\phi}+{dV(\phi)\over d\phi}-{3\over 4}{H^2\phi\over (1+\kappa^2\phi^2/8)^2}=0\,.\label{phi}
\end{equation}
From the above equations (\ref{H2}) and (\ref{Hdot}) it is easy to recognize the energy density and pressure for the Elko spinor field with torsion. These quantities are given by 
\begin{equation}
\tilde{\rho}_\phi={\dot{\phi}^2\over 2}+V(\phi)+{3\over 8}{H^2\phi^2\over (1+\kappa^2\phi^2/8)}\,,\label{rho}
\end{equation}
\begin{equation}
\tilde{p}_\phi={\dot{\phi}^2\over 2}-V(\phi)-{3\over 8}{H^2\phi^2\over (1+\kappa^2\phi^2/8)} - {1\over 4 }{\dot{H}\phi^2\over (1+\kappa^2\phi^2/8)}-{1\over 2}{H\phi\dot{\phi}\over (1+\kappa^2\phi^2/8)^2}\,.\label{pphi}
\end{equation}

Several interesting aspects can be noticed in these equations. First, the Euler-Lagrange method used to obtain the Friedmann equations by variation with respect to the lapse function $N(t)$ and the scale factor $a(t)$ leads directly to the expressions of energy density and pressure, in contrast to the standard method of introducing an energy-momentum tensor containing terms of spin connections and take its variation with respect to the metric, as done in \cite{BOE6}. In \cite{BOE6} the components of the energy-momentum tensor are obtained in the torsionless limit, which corresponds to take $h(t)=0$ in the above equations (see Eqs. (\ref{rhoA})-(\ref{pA}) of the Appendix for the energy density and pressure in the null torsion case). Here the generalized equations are given by (\ref{rho}) and (\ref{pphi}). The above equations also allow the analysis of different energy scales for the field $\phi$, where the torsion effects are important for the Elko field. Since that $\kappa^2 = 1/M_{pl}^2$, in the limit $\phi\ll M_{pl}$ all the terms $\kappa^2\phi^2/8$ in the denominators can be neglected and the torsion free equations are recovered (see Eqs. (\ref{H2AA})-(\ref{HdotAA}) of the Appendix). Thus, torsion effects are only relevant for $\phi \gtrsim M_{pl}$. Notice also that in the limit $\phi\gg M_{pl}$ the torsion function behaves as $h(t)\approx -H(t)$.

\section{Slowly varying Elko spinor field}

A prominent application of the Elko spinor field as a candidate to dark energy in the universe reveals itself as far as we consider an almost constant and homogeneous Elko spinor field distribution in the whole universe, i. e., $\phi\approx constant$. This leads to a slowly varying condition to Elko spinor field given by
\begin{equation}
\dot{\phi}\ll H\phi\,, \qquad \ddot{\phi}\ll H\dot{\phi}\,.\label{slowly}
\end{equation}
In this limit we have the Friedmann equations (\ref{H2}) and (\ref{Hdot}) as:
\begin{equation}
H^2={\kappa^2\over 3}\bigg[V(\phi)+{3\over 8}{H^2\phi^2\over (1+\kappa^2\phi^2/8)} + \rho_m\bigg]\,,\label{H2slow}
\end{equation}
\begin{equation}
-2\dot{H}-3H^2={\kappa^2}\bigg[-V(\phi)-{3\over 8}{H^2\phi^2\over (1+\kappa^2\phi^2/8)} - {1\over 4 }{\dot{H}\phi^2\over (1+\kappa^2\phi^2/8)} + p_m \bigg]\,,\label{Hdotslow}
\end{equation}
where we have also included by hand standard terms of matter density $\rho_m$ and matter pressure $p_m$ representing the contributions of matter in the evolution.

In this limit the pressure and energy density of the Elko field are related by:
\begin{equation}
p_\phi=-\rho_\phi - {1\over 4 }{\dot{H}\phi^2\over (1+\kappa^2\phi^2/8)}\,,
\end{equation}
thus, for a slowly varying Hubble parameter, $\dot{H}\approx 0$, we have $p_\phi \approx -\rho_\phi$, an almost vacuum equation of state relation, as desired for a dark energy fluid.

Taking a potential that includes a mass term and a self interaction for the Elko spinor field, $V(\phi) = {1\over 2} m^2 \phi^2 + \lambda \phi^4$, where $\lambda$ is a dimensionless coupling constant, the Friedmann equation (\ref{H2slow}) can be written as:
\begin{equation}
H^2={\kappa^2\over 3}\rho_m + {\Lambda(t)\over 3}={\kappa^2\over 3}(\rho_m + \rho_\phi) \,,\label{e13}
\end{equation} 
with $\rho_\phi=\Lambda(t)/\kappa^2$ and
\begin{equation}
\Lambda(t)=\Lambda_* + 3\beta H^2\,,\label{e14}
\end{equation}
where
\begin{equation}
\Lambda_*\equiv \kappa^2 \bigg({1\over 2} m^2 \phi^2 + \lambda \phi^4\bigg)\,, \qquad \beta\equiv {1\over 8 }{\kappa^2\phi^2\over (1+\kappa^2\phi^2/8)}\,.\label{e15}
\end{equation}
We can see that such model is exactly analogous to having a cosmological constant term $\Lambda_*$ plus a time varying cosmic term $3\beta H^2$, with $\beta$ a dimensionless constant. Several models with time varying cosmic term are present in the literature. The specific form presented here, with a $H^2$ dependence, are motivated by renormalization group to the vacuum energy \cite{ren}. Reference \cite{sandro} discuss exact analytic solutions to this model, with phenomenological $\Lambda_*$ and $\beta$ terms. Here, such terms appear related to the Elko spinor mass, coupling constant $\lambda$ and the value of the field $\phi$.

By assuming a matter content that satisfies an equation of state of the form $p_m=\omega_m \rho_m$, the Friedmann equation can be put into the following form
\begin{equation}
H^2={1\over 3(1-\beta)} \bigg[{\kappa^2}\rho_m + {\Lambda_*}\bigg]\,,\label{H22}
\end{equation}
with a conservation equation for the matter energy density
\begin{equation}
\dot{\rho}_m + 3H (1+\omega_m)\rho_m = 0\,.\label{rhom}
\end{equation}
Such system can be solved in order to constrain the parameters with available observational data.

In terms of the density parameters, Eq. (\ref{H22})
can be written as
\begin{equation}
\left({H\over H_0}\right)^2={\Omega_{m}(1+z)^3+1-\Omega_{m}-\beta\over 1-\beta},\label{H2z}
\end{equation}
where $\Omega_{m}={8\pi G \rho_{m0}\over 3 H_0^2}$ is the current matter parameter density. As we assume spatial flatness, it relates to the Elko parameter density $\Omega_\phi$ by
\begin{equation}
\Omega_{\phi}+\Omega_{m}=1 \,,\qquad \Omega_{\phi}={\Lambda_*\over 3 H_0^2}+\beta.\label{e19}
\end{equation}
The free parameters $(\Omega_m,\beta)$ must respect the physical limits $\Omega_m\geq0.04$, which corresponds to current baryon density and $\beta\geq0$, as a negative $\beta$ would give rise to formation of ghosts in the lagrangian. Analogously, it corresponds to a prohibition of vacuum creation, as required by thermodynamics.

\section{Observational Constraints}
In order to constrain this class of models, we choose to use distance luminosity from magnitudes of SNs Ia. Firstly responsible for indicate the Universe acceleration \cite{SN}, today we count with 580 SNs from Union 2.1 sample \cite{SuzukiEtAl12}. We use the SN distance modulus to constrain the model free parameters:
\begin{equation}
 \mu(z)=5\log d_L(z)+25,
\end{equation}
where $d_L(z)$ is SN distance luminosity in Megaparsecs (Mpc) at redshift $z$. The luminosity distance is given in terms of Hubble parameter as
\begin{equation}
 d_L(z)=c\int_0^z\frac{1}{H(z')}dz'=\frac{c}{H_0}\int_0^z\frac{1}{E(z')}dz',
\end{equation}
where $E(z)\equiv\frac{H(z)}{H_0}=\sqrt{{\Omega_{m}(1+z)^3+1-\Omega_{m}-\beta\over 1-\beta}}$ is a function independent of the Hubble constant. With the distance modulus available for all 580 SNs from Union, we may constrain the model free parameters ($H_0,\Omega_m,\beta$) through a $\chi^2$ statistics
\begin{equation}
 \chi^2=\sum_{i=1}^{580}\left[\frac{\mu_{o,i}-\mu(z_i,H_0,\Omega_m,\beta)}{\sigma_{\mu_i}}\right]^2,
\end{equation}
where $\mu_{obs,i}$ is the estimated distance modulus for each SN, $\sigma_{\mu_i}$ is its estimated uncertainty and $\mu(z_i,H_0,\Omega_m,\beta)$ is model predicted distance modulus at redshift $z_i$.

\begin{figure}[t]
\centerline{ 
\epsfig{figure=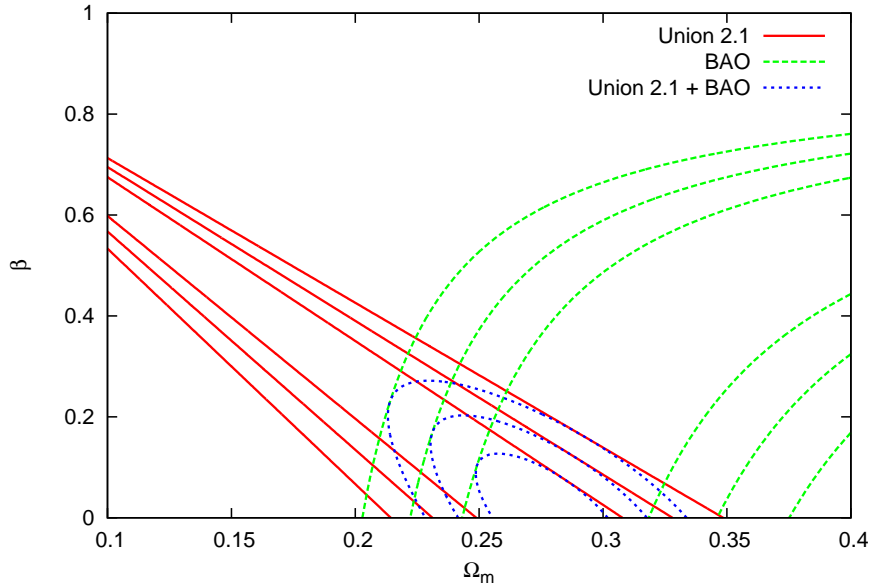,width=0.70\linewidth,angle=0}
}
\caption{Confidence contours of Elko from Union 2.1 SN data + BAO. Red, solid contours indicate SN constraints. Green, long-dashed lines indicate constraints from BAO. Blue, short-dashed lines indicate SNs + BAO constraints. The regions correspond to 68.3\%, 95.4\% and 99.7\% c.l.}
\label{contours}
\end{figure}

As usual on this type of analysis, we marginalize over the $H_0$ dependence by rewriting the distance modulus
\begin{equation}
 \mu(z)=5\log D_L(z)+M_*,
\end{equation}
where $D_L\equiv\frac{H_0d_L}{c}$ is the dimensionless luminosity distance and $M_*\equiv25+5\log \frac{c}{H_0}$ comprises all the dependence over $H_0$. We, then, marginalize the likelihood over $M_*$:
\begin{equation}
 \tilde{\mathcal{L}}(\Omega_m,\beta)=N\int_{-\infty}^{+\infty}\exp\left[-\frac{1}{2}\chi^2(M_*,\Omega_m,\beta)\right]dM_*,
\end{equation}
where $N$ is a normalization constant. The corresponding $\tilde{\chi}^2=-2\ln\left(\frac{\tilde{\mathcal{L}}}{N}\right)$ is given by
\begin{equation}
 \tilde{\chi}^2=C-\frac{B^2}{A},
\end{equation}
where
\begin{equation}
A=\sum_{i=1}^{580}\frac{1}{\sigma_i^2},\quad B=\sum_{i=1}^{580}\frac{5\log[D_L(z_i)]-\mu_{o,i}}{\sigma_i^2},\quad C=\sum_{i=1}^{580}\left\{\frac{5\log[D_L(z_i)]-\mu_{o,i}}{\sigma_i}\right\}^2. 
\end{equation}
The result of this analysis can be seen on Fig. \ref{contours}.

On this figure, we may see from the solid contours that SN data alone is not sufficient to constrain the parameter space, as there is a large degeneracy between $\Omega_m$ and $\beta$. In order to break this degeneracy, we have used the Baryon Acoustic Oscillations, which yields orthogonal constraints to luminosity distances. We have used the quantity $\mathcal{A}$ from Eisenstein {\it et al.} \cite{bao05}, defined by:
 \be
 \mathcal{A}=\frac{\Omega_m^{1/2}}{E(z_*)^{1/3}}\left(\frac{1}{z_*}\int_0^{z_*}{\frac{1}{E(z)}}\right)^{2/3},
 \ee
 where $E(z)\equiv\frac{H(z)}{H_0}$ and $z_*=0.35$ is a typical redshift from SDSS. With this definition and given the constraint \cite{bao05} of $\mathcal{A}=0.469\pm0.017$, we have added the $\chi^2$ from SN and BAO to find a combined $\chi^2_{comb}$:
 \be
 \chi^2_{comb}=\tilde{\chi}^2+\left(\frac{\mathcal{A}-0.469}{0.017}\right)^2,
 \ee
 which is equivalent to consider a Gaussian prior from BAO.

 From this analysis, we find $\Omega_m=0.278^{+0.024+0.040+0.056}_{-0.029-0.048-0.065}$, $\beta=0.00^{+0.13+0.20+0.27}_{-0.00-0.00-0.00}$, $\chi^2_\nu=0.973$, which indicates a good fitting.

\section{Linear evolution of density perturbations}
In this section we study the density perturbations related to this model, since it is well known that their effects are the responsible for large scale structure formation in the universe. The most rigorous way to treat perturbations in the cosmological context is by means of the general relativity, where the metric, the energy-momentum tensor and the fields are perturbed. In the present context also the torsion terms should be perturbed. Such an approach, however, is very difficult to apply in our case, and it is also important to remember that our equations are already in the slowly varying regime (\ref{slowly}).  Therefore, in order to address the perturbations effects in an effective way, we adopt the method suggested by Abramo {\it et al.} \cite{abramo}, where structure formation in the presence of dark energy perturbations are considered in the so called pseudo-Newtonian approach. Moreover, the model developed in \cite{abramo} applies when dark energy can be described by some parametrization for the equation of state (EoS) as a function of redshift, $\omega_\phi(z)$, being not necessary to know the specific form of the fields involved. This is the best method to be applied to our problem in order to have trustful information about the density perturbations.

In order to obtain the dark energy EoS to apply the method of Abramo {\it et al.}, notice that in our model the fluids are separately conserved, having no interaction between them. Thus, besides the matter density conservation (\ref{rhom}) we also have a dark energy density conservation,
\begin{equation}
\dot{\rho}_\phi + 3H (\rho_\phi + p_\phi) = 0\,.\label{rhophi}
\end{equation}
Assuming an EoS parameter in terms of the redshift defined by $p_\phi = \omega_\phi(z) \rho_\phi$, the above equation can be recast as:
\begin{equation}
\frac{d\rho_\phi}{dz} = \frac{3(1+\omega_\phi(z))}{(1+z)}\rho_\phi\,,\label{rhoz}
\end{equation}
where we have used $dt/dz=-1/(H(z)(1+z))$. Taking $\rho_\phi$ from (\ref{e13}), using (\ref{e14}), (\ref{e15}), (\ref{H2z}) and (\ref{e19}) we have
\begin{equation}
\rho_\phi=\frac{3H_0^2}{\kappa^2}\bigg[\frac{1-\beta-\Omega_m+\beta \Omega_m (1+z)^3}{1-\beta}\bigg]\,.\label{rhophiz}
\end{equation}
By means of Eq. (\ref{rhoz}) we finally obtain
\begin{equation}
\omega_\phi(z)=\frac{\beta - 1 +\Omega_m}{1-\beta - \Omega_m +\beta \Omega_m (1+z)^3}\,.\label{omegaz}
\end{equation}
In the limit $\beta\to0$ we recover exactly the $\Lambda$CDM model, namely $\omega_\phi=-1$, a vacuum equation of state parameter.

In order to proceed, let us introduce the cosmological perturbations by admitting inhomogeneous deviations from the background quantities in the usual way,
\begin{equation}
\rho_m=\bar{\rho}_{m}(1+\delta_m),\quad p_m=\bar{p}_m+\delta p_m,\quad\rho_\phi=\bar{\rho}_{\phi}(1+\delta_\phi),\quad p_\phi=\bar{p}_\phi+\delta p_\phi,
\end{equation}
and therefore $\delta_m=\delta\rho_m(\vec{x},t)/\bar{\rho}_{m}$ and $\delta_\phi=\delta\rho_\phi(\vec{x},t)/\bar{\rho}_{\phi}$ are the so called density contrasts for each fluid. In \cite{abramo} the non-linear differential equations characterizing the growth of spherically symmetric perturbations for arbitrary time-dependent equations of state were obtained. It was also shown that the pseudo-Newtonian formalism agree with the general relativity one for pressureless fluids.

Here we just analyze the linear regime of cosmological perturbations due to a dark energy fluid mimicked by the Elko field, which is valid for radiation and also for matter era. For this we just need the EoS parameter $\omega_\phi(z)$, since that the matter part is taken as a pressureless fluid, $\omega_m=0$. We strongly recommend the reader to the reference \cite{abramo}, where both the linear and non-linear equations for density perturbations are derived carefully. The linear regime is described by the coupled differential equations (see Eqs. (29) and (30) of \cite{abramo}):
\begin{equation}
\ddot{\delta}_m+2H\dot{\delta}_m=\frac{3H^2}{2}[\Omega_m\delta_m + \Omega_\phi \delta_\phi (1+3\omega_\phi)]\,,\label{deltam}
\end{equation}
\begin{equation}
\ddot{\delta}_\phi+\bigg(2H-\frac{\dot{\omega}_\phi}{1+\omega_\phi}\bigg)\dot{\delta}_\phi=\frac{3H^2}{2}(1+\omega_\phi)[\Omega_m\delta_m + \Omega_\phi \delta_\phi (1+3\omega_\phi)]\,.\label{deltaphi}
\end{equation}
where a dot stands for time derivative. This system of equations must be solved in order to study the growth of matter perturbation and dark energy perturbation, due to the presence of a dark energy fluid  mimicked by the Elko field by means of the EoS parameter (\ref{omegaz}). It is interesting to realize that, although DE and DM are separately conserved at background level, they are coupled at first order of density perturbations, which can lead to interesting effects and can also be used as an independent check of cosmological models. Notice that the presence of the $\beta$ parameter into $\omega_\phi$ carries explicit information that comes from the Elko field, in an effective way.

\begin{figure}[t]
\centerline{ 
\epsfig{figure=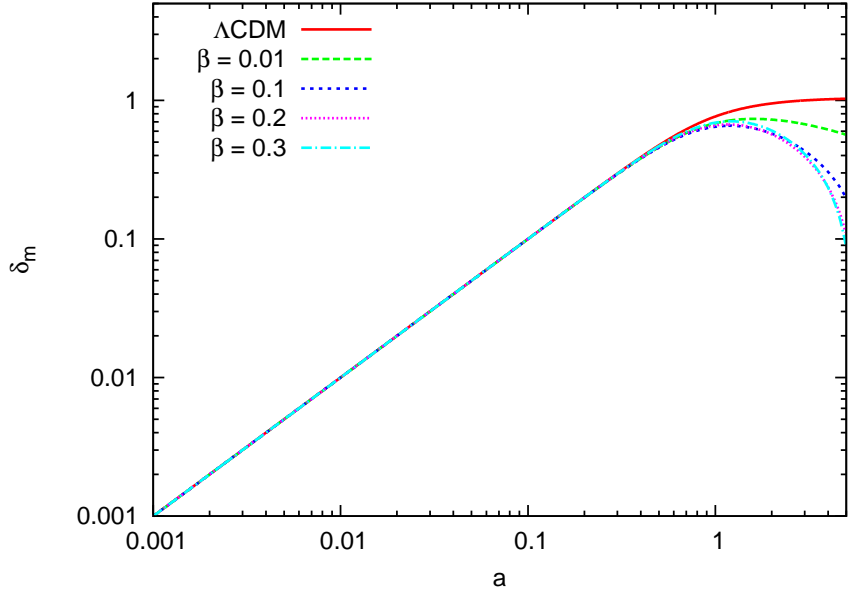,width=0.70\linewidth,angle=0}
}
\caption{Matter contrast density $\delta_m$ as a function of the scale factor for some values of the free parameter $\beta$. The $\Lambda$CDM result is also shown for comparison.}
\label{figdeltam}
\end{figure}

\begin{figure}[t]
\centerline{ 
\epsfig{figure=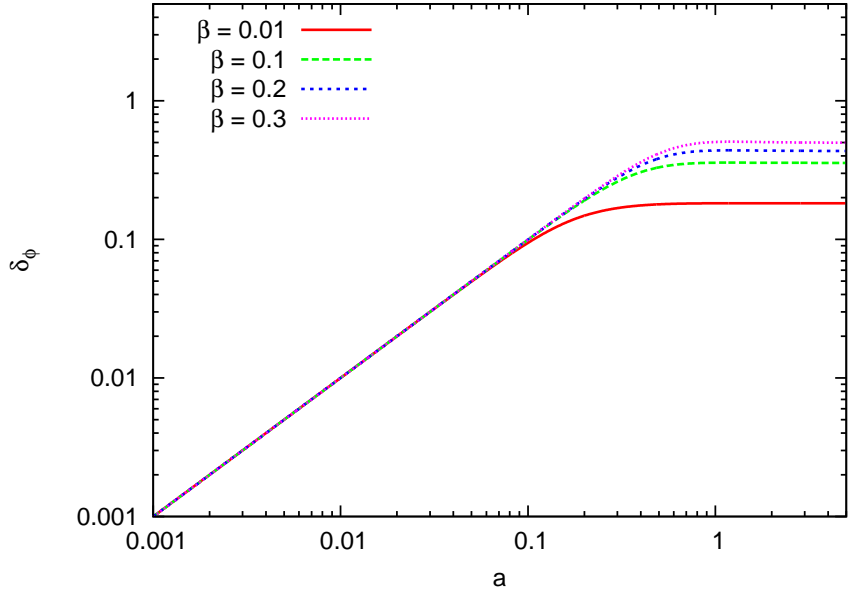,width=0.70\linewidth,angle=0}
}
\caption{Elko contrast density $\delta_\phi$ as function of the scale factor for some values of the free parameter $\beta$. $\Lambda$CDM has no dark energy perturbation.}
\label{figdeltaphi}
\end{figure}

In order to solve numerically Eqs. (\ref{deltam})-(\ref{deltaphi}), we used as initial conditions for DM the expected for Einstein-deSitter universe, namely, $\delta_m\propto a$, as DE is negligible at matter era, when we start to integrate. The main problem is the following: as DE perturbations have never been measured, we have no idea how it should behave at matter era. As we do not expect DE perturbations to grow faster than DM perturbations, a possible way to circumvent this problem is by choosing the upper limit, namely, $\delta_\phi\propto a$. As a result, we show in Fig. \ref{figdeltam} the dark matter contrast density evolution and in Fig. \ref{figdeltaphi}, the Elko contrast density evolution. As one may see on Figure \ref{figdeltam}, the dark matter perturbations in Elko cosmology are quite similar to the standard $\Lambda$CDM model, mainly up to $a=1$. In the future, $a>1$, the DM perturbations on Elko cosmology tend to decrease, however, while $\Lambda$CDM perturbations tend to be constant. Nevertheless, it must be noted that this is an approximated method, and it involves some assumptions \cite{abramo}, as $c_{eff}^2=\omega$ (i.e., $\frac{\delta p_\phi}{\delta\rho_\phi}=\omega_\phi$), for instance, in order to find an agreement between spherical collapse and neo-Newtonian formalisms (see Ref. \cite{JesusEtAl11} for other possibilities).

In Fig. \ref{figdeltaphi} we may note that the DE perturbations freeze at some point and are always smaller than DM perturbations. We have also noticed that, while from $\beta=0$ ($\Lambda$CDM) to $\beta=0.1$ there are big differences among DE perturbations growing, for $\beta>0.1$ the behaviors are quite similar. Note that for $\Lambda$CDM, there is no DE perturbation, so it is not shown on this Figure.

\begin{figure}[t]
\centerline{ 
\epsfig{figure=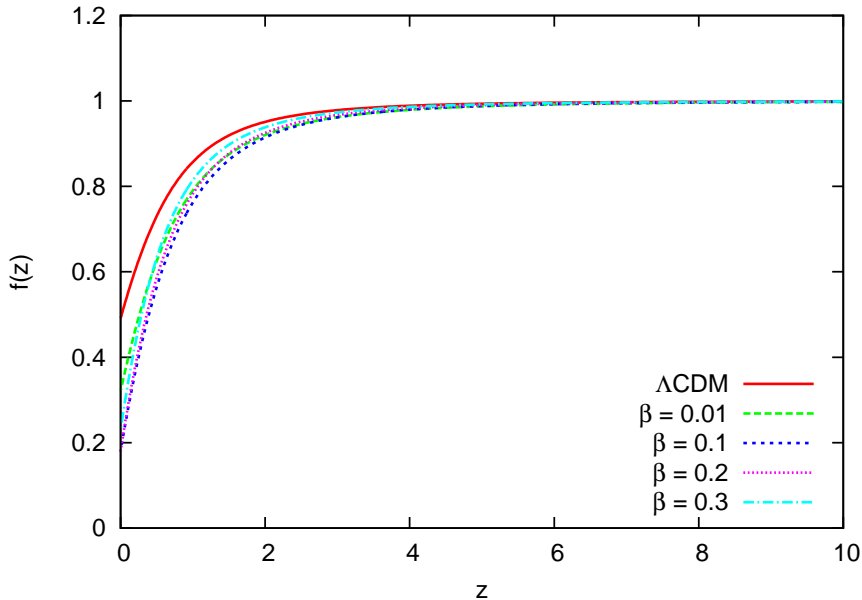,width=0.70\linewidth,angle=0}
}
\caption{Growth factor $f(z)$ as function of the redshift for some values of the free parameter $\beta$. Also shown is the $\Lambda$CDM result for comparison.}
\label{fz}
\end{figure}

It is also important to analyze the DM clustering is the growth rate \cite{Peebles}. This is an efficient parametrization of the linear matter fluctuations $\delta_m$ which has the following functional form:
\be
f(z)=\frac{d\ln\delta_m}{d\ln a}=-(1+z)\frac{d\ln\delta_m}{dz}
\ee As one may see on Fig. \ref{fz}, there is quite small differences from $\Lambda$CDM to Elko $f(z)$ at low redshifts, which may indicate a good agreement of Elko with current observations. We notice, nevertheless, that as the Elko field becomes more prominent the growth factor decreases. It may be understood by the fact that the Elko dark energy acts repelling ordinary matter, making it difficult its agglomeration.  

\section{Concluding remarks}

In this paper we have investigated a $\Lambda(t)$ cosmology of the form $\Lambda(t)=\Lambda_* + 3\beta H^2$, which follows naturally from a dark energy fluid represented by a slowly varying Elko spinor field filling the whole universe.  $\Lambda_*$ is related to the potential energy of the spinor field (\ref{e15}), here represented by a massive quadratic potential plus a self interacting quartic term. The term $3\beta H^2$ comes from the spin connection of the Elko spinor field in a FRW metric. Using (\ref{e15}), the constraints of $\beta$ and $\Omega_m$ obtained from supernovae data leads to interesting results, as follows. From (\ref{e15}) we obtain $\phi^2=8M_{pl}^2\beta/(1-\beta)\approx 8M_{pl}^2\beta$, where $M_{pl}=1/\kappa\approx 10^{19}$GeV and we have used $\beta\ll 1$ from the observational constraint, namely $\beta = 0.00^{+0.13}_{-0.00}$ at $1\sigma$. This shows that the slowly varying field $\phi$ satisfies $\phi\ll M_{pl}$, standing in the classical limit region, as expected. Another interesting result comes for the Elko mass estimate. From (\ref{e15}) and (\ref{e19}) and the above considerations for $\phi$ and $\beta$ we obtain for the Elko mass
\begin{equation}
m^2\approx\frac{3}{4}H_0^2\frac{\Omega_\phi}{\beta}-16\lambda M_{pl}^2\beta\,,\label{m2}
\end{equation}
with $\Omega_\phi=1-\Omega_m \approx 0.722$. The positivity of the mass leads to a condition on the self-coupling constant $\lambda$, namely $\lambda < \frac{3}{64}\frac{H_0^2}{M_{pl}^2}\frac{\Omega_\phi}{\beta^2}$. Although $\beta\ll 1$, the term $H_0^2/M_{pl}^2 \approx 10^{-122}$, indicates a nearly null value to coupling. It is interesting that, in a quite different context, self-coupling seems to be ruled out from unitarity issues \cite{cheng}. Even being restricted to the classical counterpart of the field, the result coming from the best fitting seems to corroborate the null value for the coupling constant. 

If we take $\lambda=0$, the Elko mass could be determined for the first term on the right side of (\ref{m2}). Although we have obtained $\beta\approx 0$ for the best fit, we have a superior limit of $0.11$ at $68\%$ c.l.. This indicates that lower the value of $\beta$ greater the value of $m$. Up to $1\sigma$ on the value of $\beta$ we have $m>2.2\times 10^{-33}$eV. The addition of other complementary observational data could constrain the $\beta$ value more accurately, leading to a better estimate for the value of the Elko mass.

In which concern the study of the evolution of contrast density of dark matter (FIG. \ref{figdeltam}) we see that smaller the value of $\beta$ the model approaches the $\Lambda$CDM model, as expected. The same behaviour is observed to the evolution of contrast density of Elko. Increasing the value of $\beta$, greater the value of $\delta_\phi$ (FIG. \ref{figdeltaphi}). The evolution of the growth factor $f(z)$ is also affected when the $\beta$ parameter increase (FIG. \ref{fz}). As mentioned, the increase of beta, encoding Elko effect, makes difficult the clustering of matter, since Elko field acts as dark energy. In all cases a subtle deviation from $\Lambda$CDM model is observed when $\beta$ increases.


\section*{APPENDIX: Elko in Einstein-Cartan framework}

The Elko spinor action in a general Einstein-Cartan framework reads
\begin{equation}
S = \int d^4 x \sqrt{-g} \left[ -\frac{1}{2\kappa^2}\tilde{R}+{1\over 2}g^{\mu\nu}\tilde{\nabla}_\mu \bar{\lambda}\tilde{\nabla}_\nu \lambda -V(\bar{\lambda}\lambda) \right] \,.
\label{action}
\end{equation}
The flat FRW metric, used throughout the paper, can be written in terms of vierbein:
\begin{equation}
g_{\mu\nu}=e_\mu^{\;\;a}e_\nu^{\;\;b}\eta_{a b},
\end{equation}
where $\eta_{a b}=diag(1,-1,-1,-1)$ and $e_\mu^{\;\;a}$ is given by
\begin{equation}
e_\mu^{\;\;a}=[N(t),a(t),a(t),a(t)]\,, \qquad e^\mu_{\;\;a}=[{1\over N(t)},{1\over a(t)},{1\over a(t)},{1\over a(t)}].
\end{equation}
Here greek indexes stands for the curved spacetime, while latin indexes are Lorentz indexes of the tangent bundle. Following the usual approach, Dirac matrices $\gamma^\mu$ in curved spacetime are related to the standard ones $\gamma^a$ by ${\gamma}^{\mu}={e^{\mu}}_{a}{\gamma}^{a}$. Obviously
\begin{equation}
\gamma^{\mu}\gamma^{\nu}+\gamma^{\nu}\gamma^{\mu}=2g^{\mu\nu}, \quad \gamma^{a}\gamma^{b}+\gamma^{b}\gamma^{a}=2\eta^{ab}.
\end{equation}

The covariant derivatives of a spinor and its dual are defined as
\begin{equation}
\tilde{\nabla}_\mu \lambda\equiv \partial_\mu\lambda - \tilde{\Gamma}_\mu \lambda\,, \qquad \tilde{\nabla}_\mu \bar{\lambda}\equiv \partial_\mu\bar{\lambda} + \bar{\lambda}\tilde{\Gamma}_\mu\,,
\label{cdd}
\end{equation}
where, as remarked in the main text, tilde denotes the presence of torsion, which is contained into spin connection ${ \tilde{\Gamma}}_{\mu}$. In order to find an expression for the spin connection we write the most general form of a covariant derivative of an object composed by flat and curved indexes as 
\begin{equation}
\tilde{\nabla}_{\lambda}{{{{W}^{a}}_{b}}^{\mu}}_{\nu}=\partial_{\lambda}{{{{W}^{a}}_{b}}^{\mu}}_{\nu}+{{\omega_{\lambda}}^{a}}_{c}{{{{W}^{c}}_{b}}^{\mu}}_{\nu}-{{\omega_{\mu}}^{c}}_{b}{{{{W}^{a}}_{c}}^{\mu}}_{\nu}+\tilde{\Gamma}_{\lambda\rho}^{\mu}{{{{W}^{a}}_{b}}^{\rho}}_{\nu}-\tilde{\Gamma}_{\lambda\nu}^{\rho}{{{{W}^{a}}_{b}}^{\mu}}_{\rho},
\label{deriv_geral}
\end{equation}
where $\tilde{\Gamma}_{\lambda\nu}^{\rho}$ is the affine connection (here including torsion terms) and ${{\omega_{\mu}}^{c}}_{b}$ its equivalent acting in the flat indexes. According to a theorem present in \cite{spinor}, we have the covariant derivative of $\gamma^\mu$ as being null, namely:
\begin{equation}
\tilde{\nabla}_{\mu}\gamma^{\nu}=\partial_{\mu}\gamma^{\nu}+\tilde{\Gamma}^{\nu}_{\mu \lambda}\gamma^{\lambda}+\left[\tilde{\Gamma}_\mu,\gamma^\nu\right]=0.
\label{deri_gamma}
\end{equation}
To discover the form of $\tilde{\Gamma}_\mu$ we also use the vierbein vanishing covariant derivative
\begin{equation}
\tilde{\nabla}_\mu{e^\nu}_a=\partial_\mu{e^\nu}_a+\tilde{\Gamma}^{\nu}_{\mu\rho} {e^\rho}_{a}-{{\omega_{\mu}}^{b}}_{a}{e^\nu}_{b}=0.
\label{der_vierbein}
\end{equation}

Finally, in order to write $\tilde{\Gamma}_\mu$ in terms of ${{\omega_{\mu}}^{c}}_{b}$, we multiply (\ref{der_vierbein}) by $\gamma^a$ and sum up with (\ref{deri_gamma}). After some calculations we find
\begin{equation}
\tilde{\Gamma}_\mu=\frac{1}{8}{\omega_{\mu}}^{ab}\left[\gamma_{a},\gamma_{b}\right]+M_\mu,
\end{equation}
where ${{\omega_{\mu}}^{ab}} = e^{a}_\nu\partial_\mu{e^\nu}^{b}+e^{a}_\nu\tilde{\Gamma}^{\nu}_{\mu\rho} {e^\rho}^{b}$ and $M_\mu$ can be set to zero for simplicity \cite{spinor}.

The affine connection containing the contorsion $K_{\;\;\mu\nu}^\rho$ is given by
\begin{equation}
\tilde{\Gamma}^\rho_{\mu\nu} = \Gamma^\rho_{\mu\nu}+K^\rho_{\;\;\mu\nu}\,,
\end{equation}
where $\Gamma^\rho_{\mu\nu}$ is the standard Christoffel symbol. The contorsion is related to the torsion tensor $T_{\;\;\mu\nu}^\rho$ via
\begin{equation}
K^\rho_{\;\;\mu\nu}=-{1\over 2}(T^\rho_{\;\;\mu\nu}+T_{\mu\nu}^{\;\;\;\;\rho}+T_{\nu\mu}^{\;\;\;\;\rho} )\,.
\end{equation}
Following a theorem presented by \cite{tsam} (see also \cite{chee} for a recent application on de Sitter solutions in quadratic gravitation), when the cosmological principle is extended to a Riemann-Cartan spacetime, we can assume the non-vanishing components of torsion as
\begin{equation}
T_{110}=T_{220}=T_{330} = -T_{101}=-T_{202}=-T_{303}= a(t)^2h(t),
\end{equation}
\begin{equation}
T_{ijk}=2a(t)^3f(t)\varepsilon_{ijk},
\end{equation}
where $a(t)$ is just a convenient factor. The functions $h(t)$ and $f(t)$ are general and $\varepsilon_{ijk}=-\varepsilon_{jik}=-\varepsilon_{ikj}$ is the totally antisymetric symbol, with $\varepsilon_{123}=1$ and $\varepsilon_{ijj}=0$. The non-vanishing components of the connection are:
\begin{equation}
\tilde{\Gamma}^0_{00}={\dot{N}\over N}\,,\quad \tilde{\Gamma}^0_{ij}={a\dot{a}+a^2h\over N^2}\delta_{ij}\,,\quad \tilde{\Gamma}^i_{0j}={\dot{a}+ah\over a}\delta_{ij}\,, \quad \tilde{\Gamma}^i_{j0}={\dot{a}\over a}\delta_{ij}\,, \quad \tilde{\Gamma}^i_{jk}=-af \varepsilon_{ijk}\,. 
\end{equation}
Therefore the curvature scalar reads
\begin{equation}
\tilde{R}=-6\Bigg[{1\over aN}{d\over dt}\bigg({\dot{a}+ah\over N}\bigg)+\bigg({ \dot{a}+ah\over aN}\bigg)^2 - {f}^2 \Bigg]\,.
\end{equation}
By assuming the Elko spinor fields as $\lambda(x^\mu)=\phi(t) \xi({\bf x})$ and $\bar{\lambda}(x^\mu)=\phi(t) \bar{\xi}({\bf x})$, such that $\bar{\xi}\xi =1$, the lagrangian density mat be recast into the form 
\begin{equation}
\mathcal{L} = -{1\over N}\bigg({3a\dot{a}^2\over \kappa^2} - {3a^3h^2\over \kappa^2} - {1\over 2}a^3\dot{\phi}^2-{3\over 8}a(\dot{a}+ah)^2\phi^2\bigg)-N\bigg({3a^3f^2\over \kappa^2}+{3\over 8}a^3f^2\phi^2+a^3V(\phi)\bigg)\,,
\end{equation}
where $V(\phi)$ is the potential.

By taking the Euler-Lagrange equations with respect to $N(t)$, $a(t)$, $\phi(t)$, $h(t)$ and $f(t)$ we obtain (setting $N\to 1$ at the end), respectively
\begin{equation}
3H^2={\kappa^2}\bigg[{\dot{\phi}^2\over 2}+V(\phi)+{3\over 8}{H^2\phi^2}+{3\over 4}{H h \phi^2}\bigg]+3\bigg(1+{1\over 8}\kappa^2\phi^2\bigg)h^2+3\bigg(1+{1\over 8}\kappa^2\phi^2\bigg){f^2} \,,\label{H2A}
\end{equation}
\begin{equation}
-2\dot{H}-3H^2={\kappa^2}\bigg[{\dot{\phi}^2\over 2}-V(\phi)-{3\over 8}{H^2\phi^2}-{1\over 4}{d\over dt}[(H+h)\phi^2]\bigg] +3\bigg(1+{1\over 8}\kappa^2\phi^2\bigg)h^2-3\bigg(1+{1\over 8}\kappa^2\phi^2\bigg){f^2}\,,\label{HdotA}
\end{equation}
\begin{equation}
\ddot{\phi}+3H\dot{\phi}+{dV(\phi)\over d\phi}-{3\over 4}\bigg((H+h)^2-{f^2}\bigg)\phi=0\,,\label{eqphiA}
\end{equation}
\begin{equation}
h(t)=-{1\over 8}{\kappa^2\phi^2\over (1+\kappa^2\phi^2/8)}\bigg({\dot{a}\over a}\bigg)\;, \qquad f(t) = 0\,,\label{hfA}\\
\end{equation}
where $H=\dot{a}/a$, as usual. Substituting $h(t)$ and $f(t)$ from (\ref{hfA}) into (\ref{H2A}), (\ref{HdotA}) and (\ref{eqphiA}) we obtain, after some algebraic manipulations, the equations (\ref{H2}), (\ref{Hdot}) and (\ref{phi}).

For the torsion free case ($h=f=0$) we obtain
 \begin{equation}
3H^2={\kappa^2}\rho_\phi\,,\label{H2AA}
\end{equation}
\begin{equation}
-2\dot{H}-3H^2={\kappa^2}p_\phi\\,,\label{HdotAA}
\end{equation}
with
\begin{equation}
\rho_\phi={\dot{\phi}^2\over 2}+V(\phi)+{3\over 8}{H^2\phi^2}\,,\label{rhoA}
\end{equation}
\begin{equation}
p_\phi={\dot{\phi}^2\over 2}-V(\phi)-{3\over 8}{H^2\phi^2}-{1\over 4}\dot{H}\phi^2 -{1\over 2}H\phi\dot{\phi}\,,\label{pA}
\end{equation}
which are, as expected, exactly the same expressions for energy density and pressure obtained in \citep{BOE6} by the standard approach.


\begin{acknowledgements}
SHP is grateful to CNPq - Conselho Nacional de Desenvolvimento Cient\'ifico e Tecnol\'ogico, Brazilian research agency, for financial support, grants number 304297/2015-1. APSS thanks to CAPES - Coordenac\~ao de Aperfeicoamento de Pessoal de N\'ivel Superior,
for financial support. JMHS thanks to CNPq (445385/2014-6;304629/2015-4) for financial support. 
\end{acknowledgements}


\end{document}